\documentclass[aps,prl,onecolumn,amsmath,amssymb,superscriptaddress,floatfix,12pt]{revtex4}

\usepackage{multirow}
\usepackage{xcolor}
\usepackage{bm}
\usepackage{amsfonts,amssymb,amsmath}
\usepackage{graphicx,dcolumn,bm,color,braket,slashed}
\usepackage{times} 
\usepackage{comment}
\usepackage{array}
\usepackage{textcomp}
\usepackage{siunitx}
\usepackage{float}

\renewcommand{\v}[1]{{\bf #1}}

\begin{document}
\title{Enhancing van-Hove singularities in SrRuO$_3$ films by vacancy engineerings}

\author{Moon Hyoung Lee}
\affiliation{Department of Physics, Ajou University, Suwon, 16499, Republic of Korea}

\author{Hyungwoo Lee}
\email{hyungwoo@ajou.ac.kr}
\affiliation{Department of Physics, Ajou University, Suwon, 16499, Republic of Korea}
\affiliation{Department of Energy Systems Research, Ajou University, Suwon, 16499, Republic of Korea}

\author{Jun-Won Rhim}
\email{jwrhim@ajou.ac.kr}
\affiliation{Department of Physics, Ajou University, Suwon, 16499, Republic of Korea}

\begin{abstract}
Flat bands, characterized by their localized electronic states and van Hove singularities, provide an ideal platform for exploring many-body physics.
However, transition metal oxides hosting flat bands are quite rare.
In this study, we investigate the origin of the existing nearly flat bands (NFBs) in SrRuO$_3$ thin films and demonstrate how to increase the number of them through structural modifications.
Using a tight-binding model that replicates experimental band structures, we analyze the SrRuO$_3$ monolayer, revealing the origin of its NFBs along the $x$ and $y$ directions. 
These NFBs arise from destructive interference stabilizing strip-type compact localized states.
By introducing periodic Ru-site vacancies, additional NFBs are generated, classified as partial or complete, depending on their Brillouin zone coverage. 
The compact localized states associated with these NFBs are identified, providing insight into their physical origin. 
For a 4-layer SrRuO$_3$ multilayer film, we uncover many partial NFBs along the $\Gamma$X and XM directions and reveal the distinct origin of their development.
Our findings highlight the potential of engineering flat bands in SrRuO$_3$ films, offering new opportunities for exploring correlated electronic phases and expanding the material platform for flat-band physics.
\end{abstract}

\maketitle

\clearpage

\section{Introduction}
From the single-particle electronic structure, one can infer potential many-body phenomena that may emerge when interactions are introduced into the system.
In this process, the density of states (DOS) around the Fermi energy often plays a crucial role.
For example, if the Stoner criterion ($UD(E_\mathrm{F}) > 1$) is satisfied, where $U$ is the Coulomb interaction strength and $D(E_\mathrm{F})$ is the DOS at the Fermi level, a ferromagnetic ground state is expected~\cite{stoner1938collective}.
Similarly, the Fermi surface nesting, enhanced when DOS contributes to the nesting increases, has often been considered one of the key indicators of charge density wave formation~\cite{chuang2001fermi,kordyuk2015pseudogap} although not all charge density waves can be explained by this mechanism~\cite{johannes2008fermi}.
In many cases, increasing the DOS is beneficial for stabilizing many-body phases, which is why flat bands featuring maximal van Hove singularities are regarded as an ideal platform for exploring many-body physics~\cite{liu2012fractional,mielke1993ferromagnetism,tasaki1998nagaoka,hase2018possibility,sharpe2019emergent,saito2021hofstadter,aoki2020theoretical,volovik1994fermi,volovik2018graphite,peri2021fragile,wu2007flat,neupert2011fractional,regnault2011fractional}.
Twisted bilayer graphene is one of the most prominent examples where superconductivity induced by flat bands has been experimentally observed~\cite{cao2018unconventional}.

The flat band exhibits a dispersion with a zero group velocity.
Electrons can be perfectly localized due to the destructive interference~\cite{bergman2008band,leykam2018artificial}.
The compact localized state (CLS) is the characteristic localized eigenstate of the flat band, which has nonzero amplitude only inside a finite spatial region~\cite{rhim2019classification,park2024quasi}.
Flat bands are classified as singular or nonsingular based on whether the corresponding Bloch wave function is continuous in momentum space~\cite{rhim2021singular}. 
In singular flat bands, CLSs alone do not span the entire flat band, necessitating the inclusion of extended states such as non-contractible loop states~\cite{bergman2008band,rhim2019classification,ma2020direct}.
Moreover, the singular flat band always touches with another dispersive band in the Brillouin zone, and the band-crossing point is characterized geometrically by the quantum distance, which is the quantum mechanical distance measuring the resemblance between two wave functions~\cite{buvzek1996quantum,dodonov2000hilbert,rhim2020quantum}.
The quantum distance governs the Landau level structures~\cite{rhim2020quantum}, transport properties~\cite{oh2024thermoelectric}, and the bulk-boundary correspondence~\cite{oh2022bulk,kim2023general} of the singular flat band.

After discovering the magic angle twisted bilayer graphene, various materials hosting flat bands have been synthesized~\cite{yin2022topological,han2021evidence,xu2020electronic,li2021dirac,yin2019negative,sun2022observation,zhang2023topological,pan2023growth,huang2024observation,fleurence2020emergence}, such as CoSn~\cite{kang2020topological}, FeSn~\cite{kang2020dirac}, and Ag/Si(111)~\cite{lee2024atomically}.
Most of them possess kagome-like lattice structures so that electronic wave functions can be localized via destructive interference.
However, despite the long history and broad scope of transition metal oxide (TMO) research~\cite{kung1989transition,tokura2000orbital,poizot2000nano,sawa2008resistive,cox2010transition,maekawa2013physics,kalantar2016two,yuan2014mixed,meyer2012transition}, experimental realizations of TMO flat band materials remain limited~\cite{jin2023electron,jovic2018dirac}.
Although there are a few theoretical predictions on the appearance of flat bands in pyrochlore lattice systems, their synthesis has been elusive~\cite{hase2019flat,hase2022quasi,hase2018computational}.
The synthesis of flat-band oxides holds promise for both fundamental and applied physics, as oxides not only provide insights into complex physical phenomena but also offer the potential for a wide range of industrial and technological applications.

In this work, we demonstrate that the van Hove singularity in SrRuO$_3$ films can be significantly enhanced by introducing a periodic arrangement of vacancies at Ru sites or by increasing the film thickness. 
To achieve this, we employ a tight-binding model with band parameters that accurately reproduce the experimentally observed band structures of SrRuO$_3$ films~\cite{sohn2021sign}.
We begin by analyzing the electronic structure of a SrRuO$_3$ monolayer, identifying the origin of its nearly flat band (NFB) above the Fermi level along the XM direction (X = $(\pi,0)$, M = $(\pi,\pi)$) through the framework of destructive interference. 
Next, we show that introducing a periodic arrangement of Ru-site vacancies, resulting in a lattice structure resembling a Lieb lattice, increases the number of NFBs in the SrRuO$_3$ monolayer. 
These bands can be classified as either partial NFBs, which span a portion of the Brillouin zone, or complete NFBs, which extend across the entire Brillouin zone. 
We elucidate the origin of these nearly flat bands by identifying all CLSs associated with them.
Finally, we explore the electronic structure of 4-unit-cell (4 u.c.) SrRuO$_3$ multilayers, demonstrating the emergence of several partial NFBs along the $\Gamma$X and XM directions. 
We reveal that the NFBs along XM are originating from the atomic CLSs, obtained by turning off the intra-layer hopping processes.
%


\section{Results}

\subsection{SrRuO$_3$ monolayer without vacancies}
\begin{figure}[!t]
\includegraphics[width=0.9 \linewidth]{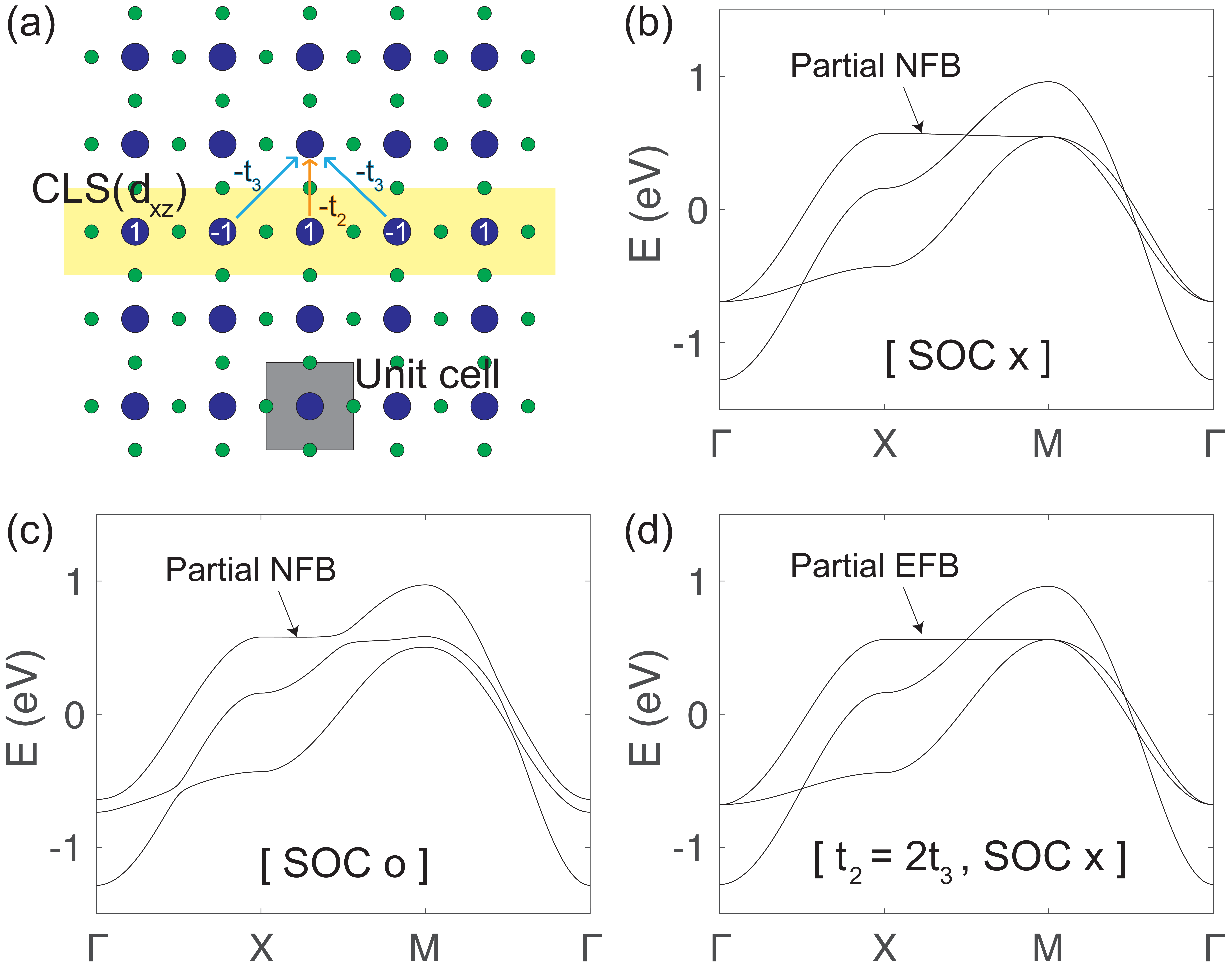}
\caption{(a) Upper view of the SRO monolayer without vacancies. Blue and green circles indicate the Ru and O atoms, respectively. Sr atoms are omitted throughout all figures, as they do not contribute significantly to the electronic structure near the Fermi level. A gray box represents the unit cell. The yellow region implies the stripe-type CLS corresponding to the partial flat band of the EFB model $H_0(\pi,k_y)$. The numbers on the blue circles mean the amplitudes of $d_{xz}$ orbitals of the CLS. In (b) and (c), we plot band dispersions of the SRO monolayer models without and with the SOC. The partial NFBs along XM of our interest are indicated by arrows. In (d), the band structure for the SRO monolayer model satisfying $t_2=2t_3$ is drawn. The arrow indicates the partial EFB along XM.}
\label{fig:Fig1}
\end{figure}

The upper view of the SrRuO$_3$ monolayer is shown in Fig.~\ref{fig:Fig1}(a), where blue and green circles represent Ru and O atoms, respectively.
We use the tight-binding Hamiltonian of the SrRuO$_3$ monolayer obtained in the previous work, where they extracted the tight-binding parameters by fitting the tight-binding band structure with the experimental one~\cite{sohn2021sign}.
They showed that the t$_{2g}$ orbitals ($d_{xy}$, $d_{yz}$, $d_{xz}$) are dominant around the Fermi level, and the Hamiltonian is given by
\begin{align}
    H = \sum_{\mathbf{k}}\sum_{m,n}\sum_{\sigma,\sigma^\prime}[\epsilon_{\mathbf{k}\sigma}^l\delta_{lm}\delta_{\sigma \sigma'} + f_{\mathbf{k}}^{lm}\delta_{\sigma \sigma'} + i\lambda\epsilon^{lmn}\tau_{\sigma \sigma'}^n]d_{\mathbf{k}l\sigma}^\dagger d_{\mathbf{k}m\sigma^\prime},\label{eq:ham_mono}
\end{align}
where indices $l$, $m$, and $n$, running from 1 to 3, represent $d$-orbitals, such that $d_{yz} \rightarrow 1$, $d_{xz} \rightarrow 2$, and $d_{xy} \rightarrow 3$.
Here, spins are denote by $\sigma$ and $\sigma'$ and $\delta_{lm}$ and $\epsilon^{lmn}$ are the Kronecker delta and Levi-Civita symboles, respectively.
The matrix elements are given by
\begin{align}
    \epsilon_{\mathbf{k}\sigma}^{1} &= -2t_{1}\cos{k_{y}} -2t_{2}\cos{k_{x}} -4t_{3}\cos{k_{x}}\cos{k_{y}},\label{eq:e_mono_1}\\
    \epsilon_{\mathbf{k}\sigma}^{2} &= -2t_{1}\cos{k_{x}} -2t_{2}\cos{k_{y}} -4t_{3}\cos{k_{x}}\cos{k_{y}},\label{eq:e_mono_2}\\
    \epsilon_{\mathbf{k}\sigma}^{3} &= -2t_{1}(\cos{k_{x}}+\cos{k_{y}}) -4t_{4}\cos{k_{x}}\cos{k_{y}},\label{eq:e_mono_3}
\end{align}
and
\begin{align}
    f_{\mathbf{k}}^{12} = f_{\mathbf{k}}^{21} = -4f\sin{k_{x}}\sin{k_{y}}.\label{eq:f_element}
\end{align}
The matrix elements of $f_{\mathbf{k}}^{lm}$ other than (\ref{eq:f_element}) are zero.
Here, (\ref{eq:f_element}) denotes the hopping between $d_{yz}$ and $d_{xz}$ orbitals at two next-nearest-neighbor Ru atoms.
The third term in (\ref{eq:ham_mono}) is the spin-orbit coupling (SOC), where $\lambda$ is the strength of the SOC and $\tau_{\sigma \sigma'}^n$ is the Pauli matrix, such that
\begin{align}
    \tau^1 = \begin{pmatrix} 0 & 1\\ 1 & 0\end{pmatrix},\quad \tau^2 = \begin{pmatrix} 0 & -i\\ i & 0\end{pmatrix},\quad\mathrm{and}\quad \tau^3 = \begin{pmatrix} 1 & 0\\0& -1\end{pmatrix}.
\end{align}
The nearest-neighbor parameters are given by $t_1=0.28$ and $t_2=0.03$ while the next nearest-neighbor hopping ones are provided by $t_3=0.018$, $t_4=0.04$, and $f=0.015$, respectively in eV.
The SOC strength is given by $\lambda=0.045$ eV.

We first revisit the electronic structures of a clean SrRuO$_3$ monolayer without vacancies.
The band dispersion when the SOC is turned off is drawn in Fig.~\ref{fig:Fig1}(b).
Each band is doubly degenerate, corresponding to spin-up and spin-down states.
Namely, there are a total of six bands.
There are many band-crossings protected by mirror and C$_4$ symmetries~\cite{sohn2021sign}.
Notably, an NFB exists along XM near $E=0.5$ eV.
Although the band-crossings are all gapped out when the SOC is turned on, the NFB remains nearly flat, as shown in Fig.~\ref{fig:Fig1}(c).
The origin of the nearly flat band (NFB) can be understood by noting that the current tight-binding model is derived by slightly modifying the band parameters of another model that hosts an exactly flat band (EFB) with exactly zero group velocity.
The EFB model for the SrRuO$_3$ monolayer is obtained by setting $t_2=2t_3$.
In this model, the EFB is developed at $E=2t_1$ along XM, as shown in Fig.~\ref{fig:Fig1}(d) for $t_2=0.3$ eV.
This EFB is a partially flat band in the sense that it is flat only along a specific direction.
Since $t_2=0.03$ and $t_3=0.018$ in the realistic model, one can consider that this model is obtained by adding a perturbation to the EFB model satisfying $t_2=2t_3$.
As a result, the EFB is slightly deformed to the NFB of the realistic SrRuO$_3$ monolayer model.
The origin of the EFB in the model with $t_2=2t_3$ can be understood from the existence of a stripe-type CLS due to destructive interference.
Along XM, the Hamiltonian is given by $H(\pi,k_y)$, and we denote the Hamiltonian for $t_2=2t_3$ by $H_0(\pi,k_y)$, which can be considered an effective 1D Hamiltonian hosting an EFB.
At the energy of the partial EFB, the Bloch eigenvector with spin-$\sigma$ is given by
\begin{align}
    | \mathbf{v}_{\mathrm{mono},\sigma}(\pi,k_y) \rangle = | d_{xz},(\pi,k_y),\sigma\rangle,\label{eq:eigvec_1}
\end{align}
where
\begin{align}
    | \xi,\mathbf{k},\sigma\rangle = \frac{1}{\sqrt{N}}\sum_{\mathbf{R}} e^{i\mathbf{k}\cdot\mathbf{R}} | \xi,\mathbf{R},\sigma\rangle,
\end{align}
is the Bloch basis vector for the $\xi$-orbital.
Here, $\sigma$ and $\mathbf{R}$ denote spins and lattice vectors.
According to previous work, a CLS can be derived from the Bloch eigenvector if its components form a finite sum of Bloch exponential factors ($e^{i\mathbf{k}\cdot\mathbf{R}}$).
For example, if the coefficient of $|\xi,\mathbf{k}\rangle$ of a flat band's Bloch eigenvector is given by $\alpha e^{-i\mathbf{k}\cdot\mathbf{R}_1}+\beta e^{-i\mathbf{k}\cdot\mathbf{R}_1}$, the corresponding CLS amplitudes for the $\xi$ orbitals in the unit cells at $\mathbf{R}_1$ and $\mathbf{R}_2$ are $\alpha$ and $\beta$, respectively.
The Bloch eigenvector (\ref{eq:eigvec_1}) satisfies this condition because the coefficient of $| d_{xz},(\pi,k_y),\sigma\rangle$ is $1 = e^{i\mathbf{k}\cdot\mathbf{R}}|_{\mathbf{R}=(0,0)}$.
Then, the coefficient of this Bloch phase 1, equal to 1, corresponds to the amplitude of the $d_{xz}$ orbital in the unit cell at $\mathbf{R}=(0,0)$, which defines the CLS being analyzed.
Considering the $y$-directional effective Hamiltonian $H_0(\pi,k_y)$, the CLS is compactly localized along the $y$-axis, with nonzero amplitude confined to the unit cell at $\mathbf{R}=(0,0)$.
Since $k_x = \pi$ is already incorporated into the effective Hamiltonian $H_0(\pi,k_y)$, the CLS extends along the $x$-axis with staggered signs, as illustrated in Fig.~\ref{fig:Fig1}(a).
%
%
%
%
This stripe-type CLS can be stabilized as an eigenmode due to destructive interference, as shown in Fig.~\ref{fig:Fig1}(a).
For example, the $d_{xz}$ orbital at the B site canceled out by two other $d_{xz}$ orbitals at the neighboring A and C sites after hopping to the D site because $t_2=2t_3$.
Although the $d_{xz}$ orbital can hop to the next nearest neighboring sites along diagonal directions, this process cannot leak the amplitude of the CLS because the sign of the corresponding overlap integral flips whenever the hopping angle changes by $\pi/2$ so that destructive interference occurs.
However, if $t_2 \neq 2t_3$ or if spin-orbit coupling (SOC) is nonzero, destructive interference no longer occurs, and the EFB cannot form.
In the case of the SrRuO$_3$ monolayer, the condition $t_2 \neq 2t_3$ is slightly violated, and the SOC strength ($\lambda$) is much smaller than the dominant hopping amplitude ($t_1$) in $H$.
As a result, the partial EFBs are not completely destroyed but instead transform into partial NFBs.
Due to C$_4$ symmetry, we have the same partial NFBs consisting of $d_{yz}$-orbitals along $\Gamma$Y, where Y represents $\mathbf{k}=(0,\pi)$.

\begin{figure}[h]
\includegraphics[width=0.9 \linewidth]{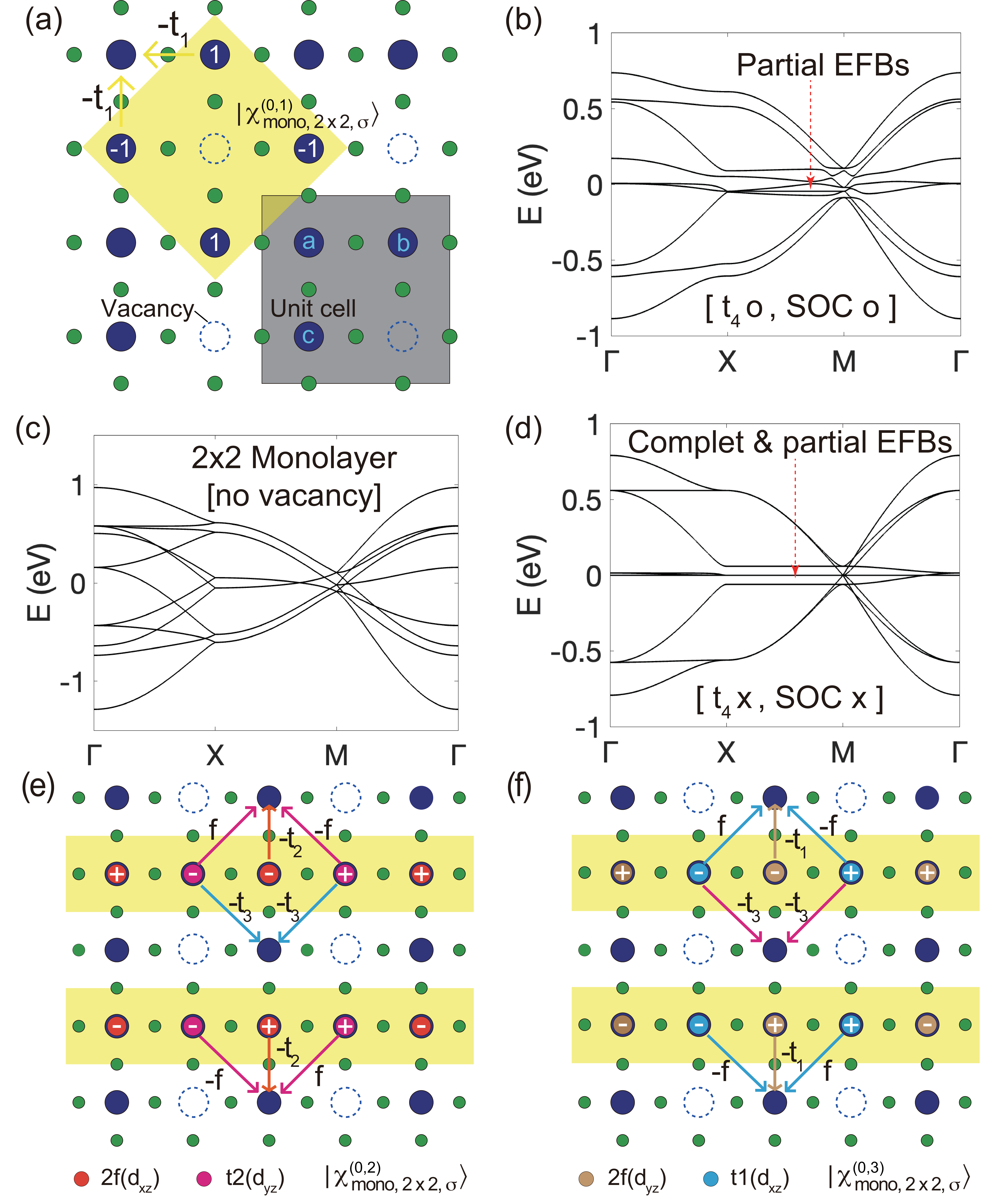}
\caption{(a) The lattice structure of the $2\times 2$ SRO monolayer. The gray box represents the unit cell, where the sublattices are denoted by a, b, and c. Vacancies are indicated by the dashed circles. The yellow region is the CLS ($|\chi^{(0,1)}_{\mathrm{mono},2\times2,\sigma}\rangle$) corresponding to the complete flat band of $H^{(0)}_{\mathrm{mono},2\times 2}(\mathbf{k})$. The yellow arrows from the b and c-sites to the neighboring a-site explain destructive interference after the hopping processes. From (b) to (d), we plot band structures for various cases where $t_4$ or SOC is either turned on or off. In (e) and (f), we draw the CLSs $|\chi^{(0,2)}_{\mathrm{mono},2\times2,\sigma}\rangle$ and $|\chi^{(0,3)}_{\mathrm{mono},2\times2,\sigma}\rangle$ corresponding to the zero-energy partial EFBs of $H^{(0)}_{\mathrm{mono},2\times 2}(\mathbf{k})$ along XM. The amplitudes of the $d$ orbitals in these CLSs are indicated by distinct colors and described at the bottom.}
\label{fig:Fig2}
\end{figure}

\subsection{SrRuO$_3$ monolayer with periodic arrangements of vacancies}
In this subsection, we demonstrate that additional flat bands can be created by engineering destructive interference through the periodic removal of Ru atoms.
A comparable approach was explored in the biphenylene network, where fluorinating carbon atoms was shown to be analogous to generating vacancies, enabling the formation or elimination of type-II Weyl semimetal states.
We arrange the vacancies in square lattice patterns.
Two vacancy configurations with 3 and 15 Ru atoms per unit cell are considered, denoted by $2\times2$ and $4\times4$ SRO monolayers, respectively.

First, we investigate the electronic structure of the $2\times2$ SRO monolayer.
The lattice structure is illustrated in Fig.~\ref{fig:Fig2}(a).
Let us denote the corresponding Bloch Hamiltonian by $H_{\mathrm{mono},2\times 2}(\mathbf{k})$, where all the tight-binding parameters are turned on, including the SOC.
The corresponding band dispersion is shown in Fig.~\ref{fig:Fig2}(b).
For comparison, we plot the band structure of the SrRuO$_3$ monolayer without vacancies in Fig.~\ref{fig:Fig2}(c).
%
While the partial NFBs around $E=\pm 0.6$ eV along $\Gamma$X originate from the partial NFBs of the clean SrRuO$_3$ monolayer by band folding, we further obtain emergent flat bands near zero energy.
The flat bands in the middle consist of a complete NFB spanning the entire Brillouin zone, along with a partial EFB and two partial NFBs along the XM direction.
Notably, one of the partial flat bands remains perfectly flat despite the presence of spin-orbit coupling (SOC).

\begin{figure}[h]
\includegraphics[width=0.9 \linewidth]{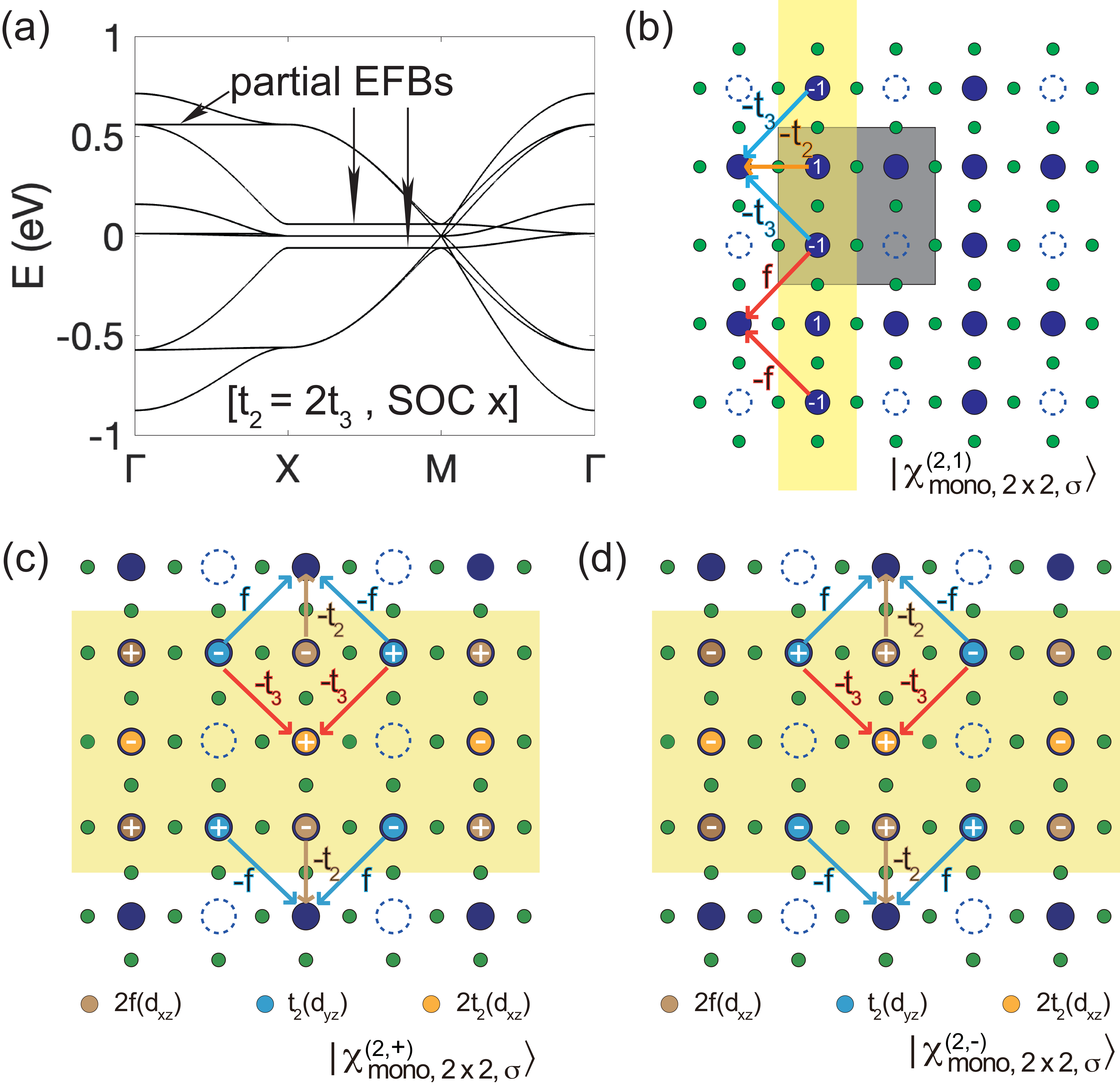}
\caption{
(a) The band structure of the $2\times 2$ SRO monolayer with $t_2=2t_3$. In this model, the SOC is turned off. In addition to the EFB at the zero energy, there are partial EFBs along $\Gamma$X and XM indicated by the arrows.
(b) The CLS corresponding to the partial EFB along $\Gamma$X, indicated by the arrow in (a), is highlighted with a yellow stripe. The numbers on the lattice sites denote the amplitudes of the $d_{yz}$-orbitals of the CLS. Hopping processes with different parameters ($-t_3$, $-t_2$, $\pm f$) are illustrated by arrows of distinct colors. In panels (c) and (d), the CLSs corresponding to the partial EFBs along the XM direction, indicated by the arrows, are shown. The amplitudes of the orbitals are provided below the figures, while their signs are indicated on the lattice sites within the CLS stripes.
}
\label{fig:Fig3}
\end{figure}

To reveal the origin of the NFBs in the middle, we analyze the electronic properties of a simplified $2\times2$ SRO monolayer model obtained by turning off $t_4$ and the SOC.
The corresponding Hamiltonian is denoted by $H^{(0)}_{\mathrm{mono},2\times 2}(\mathbf{k})$, whose band dispersion is shown in Fig.~\ref{fig:Fig2}(d).
One can consider the terms related to $t_4$ and the SOC in $H_{\mathrm{mono},2\times 2}(\mathbf{k})$ as perturbations for $H^{(0)}_{\mathrm{mono},2\times 2}(\mathbf{k})$.
In the unperturbed model, $H^{(0)}_{\mathrm{mono},2\times 2}(\mathbf{k})$, we have a complete EFB and two partial EFBs at zero energy, along with two partial EFBs at $E\approx \pm 0.05$ eV for each spin.
First, the unnormalized eigenvector corresponding to the zero energy complete EFB is given by
\begin{align}
    |\mathbf{v}^{(0,1)}_{\mathrm{mono},2\times2,\sigma}(\mathbf{k})\rangle = (e^{ik_x}+e^{i(k_x+k_y)})|d^{(b)}_{xy},\mathbf{k},\sigma \rangle + (-1-e^{ik_x})|d^{(c)}_{xy},\mathbf{k},\sigma \rangle,\label{eq:eigvec_2}
\end{align}
where $d^{(\eta)}_{xy}$, $d^{(\eta)}_{xz}$, and $d^{(\eta)}_{yz}$ are $d$-orbitals at the $\eta$-the sublattice ($\eta=a,b,c$).
From the first term of the right-hand side of (\ref{eq:eigvec_2}), one can read that the CLS amplitudes of the $d_{xy}$-orbitals at the $b$-sites in $\mathbf{R}=(-1,0)$ and $\mathbf{R}=(-1,-1)$ are 1.
On the other hand, the second term reveals that the CLS amplitudes of $d_{xy}$-orbitals at the $c$-sites in $\mathbf{R}=(0,0)$ and $\mathbf{R}=(-1,0)$ are $-1$.
Let us denote this CLS by $|\chi^{(0,1)}_{\mathrm{mono},2\times2,\sigma}\rangle$.
This CLS, depicted in Fig.~\ref{fig:Fig2}(a), functions as an eigenmode facilitated by destructive interference.
For example, the $d_{xy}$-orbitals at the $b$ and $c$ sites cancel each other after hopping to the neighboring $a$ site because they have opposite signs.
Since the next nearest neighbor hopping amplitude in this EFB model $H^{(0)}_{\mathrm{mono},2\times 2}(\mathbf{k})$ is vanishing, the CLS can be stabilized as in the case of the well-known Lieb lattice model.
On the other hand, the $d_{xz}$- and $d_{yz}$-orbitals cannot form such a Lieb-like CLS because $t_3$ corresponding to the diagonal hopping processes is nonzero.
This is why the CLS consists only of $d_{xy}$-orbitals.
However, when $t_4$ and SOC are turned on, $d_{xy}$-orbitals also cannot satisfy the condition of destructive interference, and the complete EFBs deform to the complete NFBs, as shown in Fig~\ref{fig:Fig2}(b).

In the EFB model $H^{(0)}_{\mathrm{mono},2\times 2}(\mathbf{k})$ investigated above, we have two additional partial EFBs along XM for each spin.
Namely, we have triply degenerate EFBs at $k_x=\pi$ in cluding the part of the complete EFB for each spin. 
The unnormalized Bloch eigenvectors corresponding to these two additional partial EFBs are evaluated as
\begin{align}
   | \mathbf{v}^{(0,2)}_{\mathrm{mono},2\times 2,\sigma}(\pi,k_y)\rangle = 2f(1-e^{-ik_y})|d_{xz}^{(a)},(\pi,k_y),\sigma\rangle + t_2(1+e^{-ik_y})|d_{yz}^{(b)},(\pi,k_y),\sigma\rangle,
\end{align}
and
\begin{align}
   | \mathbf{v}^{(0,3)}_{\mathrm{mono},2\times 2,\sigma}(\pi,k_y)\rangle = 2f(1-e^{-ik_y})|d_{yz}^{(a)},(\pi,k_y),\sigma\rangle + t_1(1+e^{-ik_y})|d_{xz}^{(b)},(\pi,k_y),\sigma\rangle.
\end{align}
From these two unnormalized Bloch eigenvectors, one can derive two CLSs characterizing the partial EFBs.
First, these two CLSs are stripe-type structures with alternating signs along the $x$-axis because $k_x$ is fixed at $\pi$.
Along the $y$-direction, $\mathbf{v}^{(0,2)}_{\mathrm{mono},2\times 2,\sigma}(\pi,k_y)$ reads that the corresponding CLS, denoted by $|\chi^{(0,2)}_{\mathrm{mono},2\times2,\sigma}\rangle$, has amplitudes $2f$ and $-2f$ for $d_{xz}$ orbitals in the unit cells at $\mathbf{R}=(0,0)$ and $\mathbf{R}=(0,1)$, respectively, and an amplitude $t_2$ for $d_{yz}$ orbitals in the same unit cells.
On the other hand, the CLS corresponding to $\mathbf{v}^{(0,3)}_{\mathrm{mono},2\times 2,\sigma}(\pi,k_y)$, indicated by $|\chi^{(0,3)}_{\mathrm{mono},2\times2,\sigma}\rangle$, has amplitudes $2f$ and $-2f$ for $d_{yz}$ orbitals in the unit cells at $\mathbf{R}=(0,0)$ and $\mathbf{R}=(0,1)$, respectively, and an amplitude $t_1$ for $d_{xz}$ orbitals in the same unit cells.
These two CLSs are illustrated in Fig.~\ref{fig:Fig2}(e) and (f).

Interestingly, there are doubly degenerate partial EFBs along XM even after we turn on both $t_4$ and the SOC, as indicated in Fig.~\ref{fig:Fig2}(c).
The corresponding Hamiltonian is denoted by $H^{(1)}_{\mathrm{mono},2\times 2}(\mathbf{k})$.
These partial EFBs of $H^{(1)}_{\mathrm{mono},2\times 2}(\mathbf{k})$ can be regarded as originating from the zero energy partial EFBs of $H^{(0)}_{\mathrm{mono},2\times 2}(\mathbf{k})$ because the stripe-type CLS corresponding to the partial EFB of $H^{(1)}_{\mathrm{mono},2\times 2}(\mathbf{k})$ is a linear combination of the $|\chi^{(0,2)}_{\mathrm{mono},2\times2,\sigma}\rangle$ and $|\chi^{(0,3)}_{\mathrm{mono},2\times2,\sigma}\rangle$, such that
\begin{align}
   | \chi^{(1,1)}_{\mathrm{mono},2\times 2,\sigma} \rangle = |\chi^{(0,2)}_{\mathrm{mono},2\times2,\sigma}\rangle - i|\chi^{(0,3)}_{\mathrm{mono},2\times2,\sigma}\rangle + (t_1 + t_2) \sum_{n\in \mathcal{Z}} (-1)^n\mathbf{T}(n\mathbf{a}_x)|\chi^{(0,1)}_{\mathrm{mono},2\times2,\bar{\sigma}}\rangle,\label{eq:cls_soc}
\end{align}
where $\bar{\sigma}$ represents a spin flip, $\mathbf{T}(\mathbf{r})$ is a translation operator satisfying $\mathbf{T}(\mathbf{t})f(\mathbf{r}) = f(\mathbf{r}-\mathbf{t})$, $\mathbf{a}_x$ is the primitive vector along the $x$-axis, and $\mathcal{Z}$ denotes a set of all integers.
We assume that all the CLSs used on the right-hand side of (\ref{eq:cls_soc}) are chosen so that the $y$-components of their center positions are the same.
Note that $| \chi^{(1,1)}_{\mathrm{mono},2\times 2,\sigma} \rangle$ is a zero-energy eigenmode when $t_4$ is turned on while the SOC is turned off.
Therefore, we only need to show that $| \chi^{(1,1)}_{\mathrm{mono},2\times 2,\sigma} \rangle$ is an eigenmode of the SOC part of $H^{(1)}_{\mathrm{mono},2\times 2}(\mathbf{k})$.
The SOC is an onsite operation between $d$-orbitals.
Let us consider, as an example, $d$-orbitals at A-site the upper row of the stripe-type CLS.
The initial amplitudes of $| \chi^{(1,1)}_{\mathrm{mono},2\times 2,\sigma} \rangle$ at the upper A-site are \((d_{yz,\uparrow}, d_{xz,\uparrow}, d_{xy,\downarrow}) = (-2f, 2if, 0)\). Calculating the amplitudes of each orbital after the hopping process yields the following results: 
First, \(d_{yz,\uparrow}\) at site A, hopping from \(d_{xz,\uparrow}\) results in an amplitude of \(-2i f \lambda\).  
Second, \(d_{xz,\uparrow}\) at site A, hopping from \(d_{yz,\uparrow}\) results in an amplitude of \(2f \lambda\).  
Lastly, \(d_{xy,\downarrow}\) at site A, the hopping contributions of \(d_{yz,\uparrow}\) and \(d_{xz,\uparrow}\) cancel each other out, resulting in an amplitude of 0.  
Similarly, for site B and the lower sites, the calculations show that the resulting amplitudes are proportional to \(-\lambda\) times the original CLS amplitudes.
As a result, the energy of the partial EFB becomes \(-\lambda\).


\begin{figure}[h]
\includegraphics[width=0.85\linewidth]{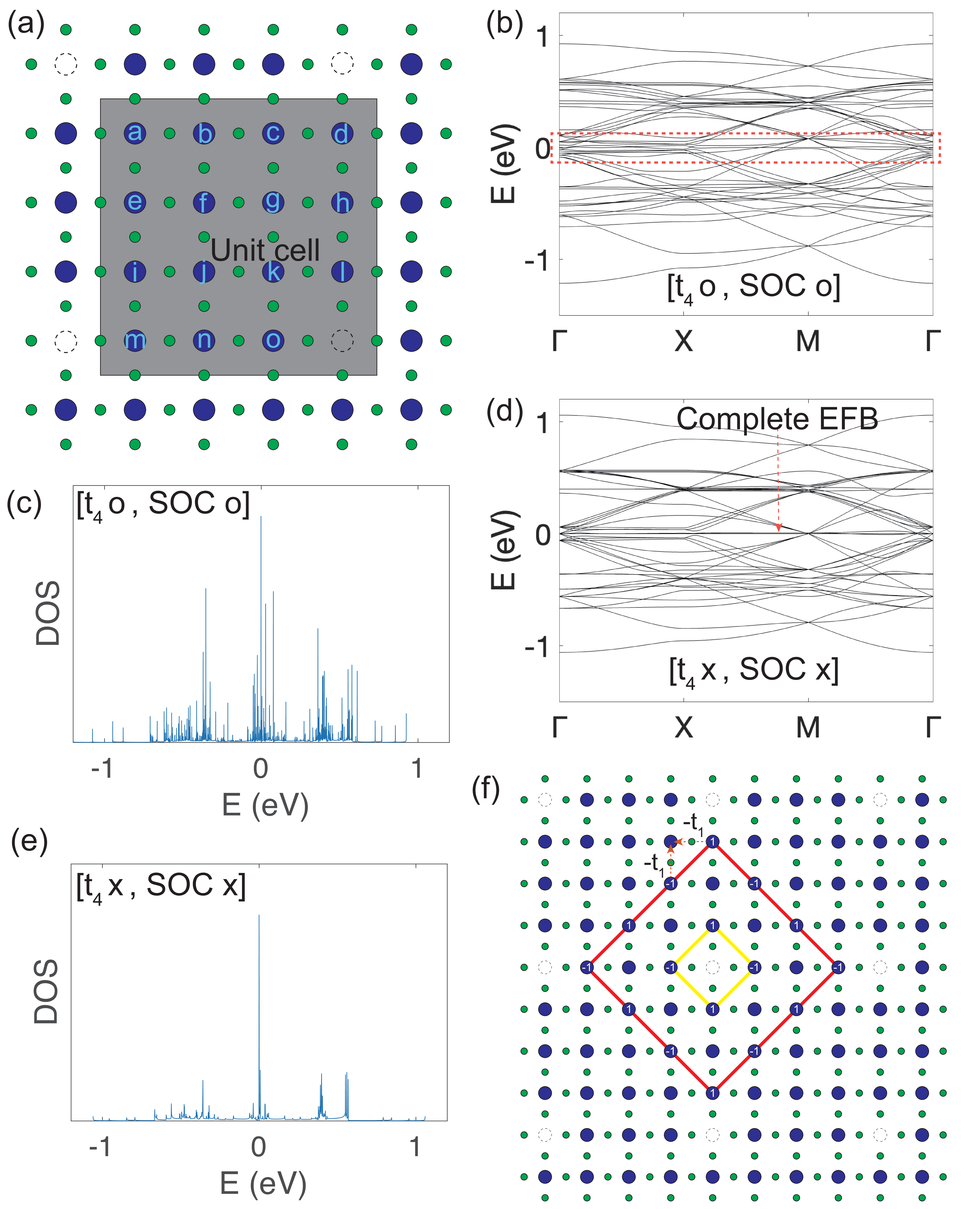}
\caption{(a) Lattice structure of the $4 \times 4$ SRO monolayer. Within the unit cell (grey region), the sublattices are labeled alphabetically. Dashed empty circles represent vacancies.
(b) Band structure with $t_4$ and spin-orbit coupling (SOC) turned on. The dashed red box highlights the nearly flat bands.
(c) Density of states (DOS) corresponding to the band structure in (b).
(d) Band structure with $t_4$ and SOC turned off. A complete exact flat band (EFB) appears at zero energy.
(e) Corresponding DOS for the band dispersion in (d). The compact localized state (CLS) associated with the EFB in (d) is illustrated. Numbers on the lattice sites indicate the $d_{xy}$-orbital amplitudes of the CLS, while solid lines serve as visual guides outlining its shape.}
\label{fig:Fig4}
\end{figure}

In addition to these complete and partial NFBs near zero energy of the full model $H_{\mathrm{mono},2\times 2}(\mathbf{k})$, we also have several partial NFBs at higher energies along $\Gamma$X ($E\approx 0.6$ eV) and XM ($E\approx \pm 0.05$ eV), as shown in Fig.~\ref{fig:Fig2}(b).
We further tune the tight-binding parameters such that $t_2=2t_3$ to find the origin of the partial NFB along $\Gamma$X in the final model.
The Hamiltonian corresponding to this parameter set is denoted by $H^{(2)}_{\mathrm{mono},2\times 2}(\mathbf{k})$.
The band dispersion for this model is plotted in Fig.~\ref{fig:Fig3}(a).
On $\Gamma$X, namely at $k_y=0$, we have a stripe-type CLS extended along the $y$-direction for the EFB at $E\approx \pm 0.6$, which is described by a unnormalized Bloch eigenvector given by
\begin{align}
   | \mathbf{v}^{(2,1)}_{\mathrm{mono},2\times2,\sigma}(k_x,0)\rangle = | d^{(a)}_{yz},(k_x,0),\sigma\rangle - | d^{(c)}_{yz},(k_x,0),\sigma\rangle.
\end{align}
This eigenvector is translated into the CLS denoted by $| \chi^{(2,1)}_{\mathrm{mono},2\times2,\sigma}\rangle$ illustrated in Fig.~\ref{fig:Fig3}(b).
As in the case of the clean SRO monolayer, the condition $t_2=2t_3$ ensures perfect destructive interference.
The eigenenergy of this CLS is $2t_1$.
On the other hand, the Bloch eigenvectors for the flat bands at $E\approx \pm 0.05$ eV are given by
\begin{align}
   | \mathbf{v}^{(2,\pm)}_{\mathrm{mono},2\times2,\sigma}(\pi,k_y)\rangle = (1+e^{ik_y})\left\{ 2f| d^{(a)}_{xz},(\pi,k_y),\sigma\rangle + t_2 | d^{(b)}_{yz},(k_x,0),\sigma\rangle \right\} \mp | d^{(c)}_{xz},(k_x,0),\sigma\rangle.
\end{align}
%
These two flat bands persist even if $t_2\neq 2t_3$ and $t_4\neq 0$.
The corresponding stripe-type CLSs, denoted by the $| \chi^{(2,\pm)}_{\mathrm{mono},2\times2,\sigma}\rangle$, are plotted in Fig.~\ref{fig:Fig3}(c) and (d).
Unlike the zero-energy CLSs along XM, these finite-energy CLSs have nonzero amplitudes for $d_{xz}$-orbitals at c-sites.
After the reflection of hopping processes, the amplitudes of $d_{xz}$-orbitals at c-sites change from $2t_2$ to $\pm8t_2f$ for $| \chi^{(2,\pm)}_{\mathrm{mono},2\times2,\sigma}\rangle$, implying that the flat band's energy is $\pm 4f$.
Note that another condition $t_2=2f$ necessary for the destructive interference is already satisfied in the initial parameter set, where $t_2=0.03$ eV and $f=0.015$ eV.
The flatness of these partial flat bands is vulnerable to the SOC, and they deform to the partial NFBs, as illustrated in Fig~\ref{fig:Fig3}(c).

%
%
%
%
%

Secondly, let us consider the $4\times4$ SRO monolayer.
In this case, we have one vacancy and 15 Ru atoms in a unit cell, as illustrated in Fig.~\ref{fig:Fig4}(a).
The band dispersion is plotted in Fig.~\ref{fig:Fig4}(b).
Among many partial and complete NFBs, we focus on the complete NFB near zero energy.
For the $2\times 2$ SRO monolayer, the complete NFB at the center was easily understood due to its resemblance to the Lieb lattice.
On the other hand, in this $4\times4$, we need to consider a completely different type of the CLS for the understanding og the NFB at the center because this system is not similar to the Lieb lattice anymore.
For the CLS analysis, we investigate an EFB model $H^{(0)}_{\mathrm{mono},4\times 4}(\mathbf{k})$ obtained by setting $t_4=0$ in the full Hamiltonian $H_{\mathrm{mono},4\times 4}(\mathbf{k})$.
Since $t_4$ is tiny in the complete model $H_{\mathrm{mono},4\times 4}(\mathbf{k})$, one can regard that the origin of the EFB of $H^{(0)}_{\mathrm{mono},4\times 4}(\mathbf{k})$ is equivalent to the origin of the NFB of $H_{\mathrm{mono},4\times 4}(\mathbf{k})$.
The EFB model $H^{(0)}_{\mathrm{mono},4\times 4}(\mathbf{k})$ yields a complete EFB at zero energy, as shown in Fig.~\ref{fig:Fig4}(c).
The corresponding eigenvector is given by
\begin{align}    |\mathbf{v}^{(0)}_{\mathrm{mono},4\times4,\sigma}\rangle =&  (e^{ik_y}+e^{i(k_x+k_y)})\ |d^{(b)}_{xy},\mathbf{k},\sigma \rangle + (1+e^{ik_x}) \left\{ 
    d^{(j)}_{xy},\mathbf{k},\sigma \rangle - d^{(m)}_{xy},\mathbf{k},\sigma \rangle - d^{(o)}_{xy},\mathbf{k},\sigma \rangle \right\} \nonumber\\
    & + (1+e^{ik_y})|d^{(e)}_{xy},\mathbf{k},\sigma \rangle + (e^{ik_x}+e^{i(k_x+k_y)}) \left\{ d^{(d)}_{xy},\mathbf{k},\sigma \rangle - d^{(g)}_{xy},\mathbf{k},\sigma \rangle + d^{(l)}_{xy},\mathbf{k},\sigma \rangle \right\},
\end{align}
which is Fourier transformed to the CLS ($| \chi^{(0)}_{\mathrm{mono},4\times4,\sigma}\rangle$), illustrated in Fig.~\ref{fig:Fig4}(b).
The CLS consists of only $d_{xy}$ orbitals.
It surrounds a vacancy, denoted by $V_0$, and its amplitude is nonzero on the sides of two $\pi/4$-rotated squares.
The four corners of the outer square touch the vacancies indicated by $V_i$ ($i>0$), and this feature is essential for stabilizing this CLS as an eigenmode because $d_{xy}$-orbitals at these corners cannot hop to these vacancy sites.
On the other hand, after the $d_{xy}$-orbitals at the sides of the squares hop to the neighboring Ru sites, we still have zero amplitudes at these neighboring sites due to destructive interference.
Due to the existence of the complete EFB, the DOS plotted in Fig.~\ref{fig:Fig4}(d) shows an exceptionally high peak at zero energy.
Generalizing the CLS analysis of $2\times 2$ and $4\times 4$ SRO monolayers, one can note the existence of a zero-energy CLS with $n$ concentric squares of nonzero amplitudes in any $2n\times 2n$ SRO monolayer.
\begin{figure}[h]
\includegraphics[width=1 \linewidth]{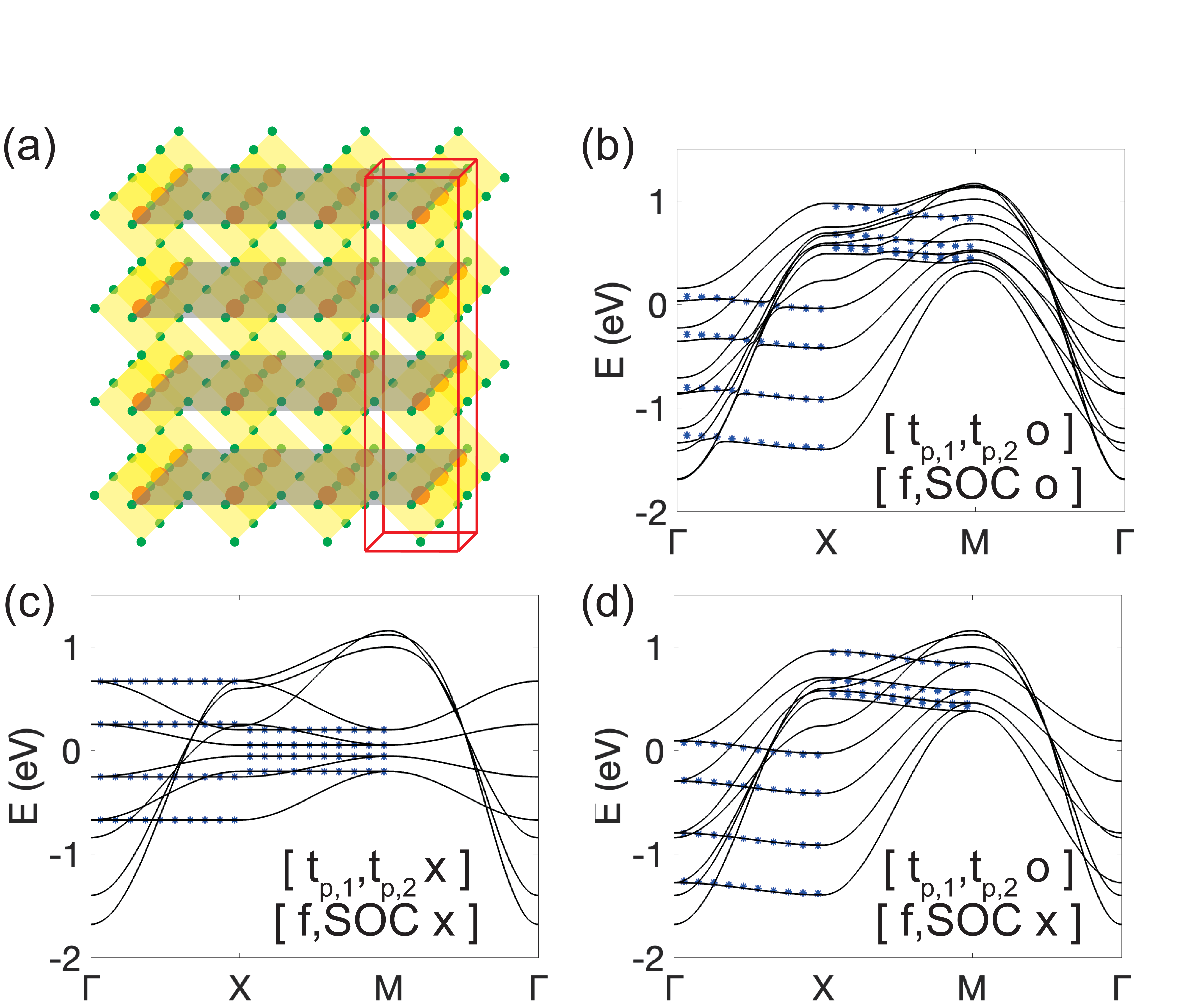}
\caption{(a) Lattice structure of the four-layer SrRuO$_3$. The red box indicates the unit cell.
We plot band dispersions of the four-layer SrRuO$_3$ for the cases where (b) all the tight-binding parameters are turned on, (c) $t_{p,1}$, $t_{p,2}$, $f$, and $\lambda$ are identically zero, and (d) only $f$, and $\lambda$ are turned off.
In (c), partial exact flat bands (EFBs) along the $\Gamma$X and XM directions are highlighted with dotted lines.
In (b) and (d), the dotted curves represent the band dispersions obtained via perturbation theory applied to the partial EFBs in (c).}
\label{fig:Fig5}
\end{figure}

\subsection{SrRuO$_3$ multilayer without vacancies}
Various SrRuO$_3$ thin films of different thicknesses have been synthesized, revealing a strong dependence of their electronic and magnetic properties on film thickness~\cite{toyota2005thickness,zhu2022thickness,ishigami2015thickness,wu2020thickness}. 
Motivated by this, we focus on a 4-layer SrRuO$_3$ film to study the van Hove singularity and explore its underlying origin.
The lattice structure and unit cell are illustrated in Fig.~\ref{fig:Fig5}(a).
The Hamiltonian of the 4-layer SrRuO$_3$ film, proposed in the previous work, is given by
\begin{align}
    H_{\mathrm{4-layer}} = H_{\mathrm{4-layer},\mathrm{intra}} + H_{\mathrm{4-layer},\mathrm{inter}},
\end{align}
where $H_\mathrm{intra}$ and $H_\mathrm{inter}$ consist of intra-layer and inter-layer interactions.
First, the intra-layer part of the Hamiltonian is written as
\begin{align}
    H_{\mathrm{4-layer},\mathrm{intra}} = \sum_{\mathbf{k}}\sum_{p}\sum_{m,n}\sum_{\sigma,\sigma^\prime}[\epsilon_{p,\mathbf{k}\sigma}^l \delta_{lm}\delta_{\sigma \sigma'} + f_{\mathbf{k}}^{lm}\delta_{\sigma \sigma'} + i\lambda\epsilon^{lmn}\tau_{\sigma \sigma'}^n]d_{p,\mathbf{k}l\sigma}^\dagger d_{p,\mathbf{k}m\sigma^\prime},\label{eq:ham_multi}
\end{align}
which is basically the same with the monolayer Hamiltonian (\ref{eq:ham_mono}) except for the layer index $p$ in (\ref{eq:ham_multi}), running from 1 to 4.
The diagonal matrix elements are given by
\begin{align}
    \epsilon_{p,\mathbf{k}\sigma}^{1} &= -2t_{p,1}\cos{k_{y}} -2t_{p,2}\cos{k_{x}},\label{eq:e_multi_1}\\
    \epsilon_{p,\mathbf{k}\sigma}^{2} &= -2t_{p,1}\cos{k_{x}} -2t_{p,2}\cos{k_{y}},\label{eq:e_multi_2}\\
    \epsilon_{p,\mathbf{k}\sigma}^3 &= U -2t_{p,3}(\cos{k_{x}}+\cos{k_{y}}) -4t_{p,4}\cos{k_{x}}\cos{k_{y}},\label{eq:e_multi_3}
\end{align}
where $\epsilon_{p,\mathbf{k}\sigma}^{1}$, $\epsilon_{p,\mathbf{k}\sigma}^{2}$, and $\epsilon_{p,\mathbf{k}\sigma}^{3}$ correspond to $d_{yz}$-, $d_{zx}$-, and $d_{xy}$-orbitals, and $U$ is the onsite energy for the $d_{xy}$-orbital.
The inter-orbital term $f_{\mathbf{k}}^{lm}$ is the same with (\ref{eq:f_element}).
The tight-binding parameters obtained in the previous work are listed in Table~\ref{table:multi_layer}.
Second, the inter-layer part of the Hamiltonian is given by
\begin{align}
    H_{\mathrm{4-layer},\mathrm{inter}} = \sum_{\mathbf{k}}\sum_{p,q}\sum_{m,n}\sum_{\sigma,\sigma^\prime}t_{qm\sigma^\prime}^{pl\sigma}(\mathbf{k}) d_{p,\mathbf{k}l\sigma}^\dagger d_{q,\mathbf{k}m\sigma^\prime},\label{eq:ham_multi_inter}
\end{align}
where
\begin{align}
    t_{pm\sigma^\prime}^{ql\sigma}(\mathbf{k}) = \delta_{p,q-1} \left[ \epsilon^l_{p,\mathbf{k}\sigma}\delta_{lm}\sigma_{\sigma\sigma^\prime} + f^{lm}_{p,\mathbf{k}}\delta_{\sigma\sigma^\prime} \right],
\end{align}
with $\epsilon^1_{p,\mathbf{k}\sigma} = -u_{p,1}-u_{p,2}\cos k_y$, $\epsilon^2_{p,\mathbf{k}\sigma} = -u_{p,1}-u_{p,2}\cos k_x$, $\epsilon^3_{p,\mathbf{k}\sigma} = -t_{p,2}$, $f^{13}_{p,\mathbf{k}} = f^{31}_{p,\mathbf{k}} = 2if\sin k_x$, and $f^{23}_{p,\mathbf{k}} = f^{32}_{p,\mathbf{k}} = 2if\sin k_y$.
Other components of $f^{lm}_{p,\mathbf{k}}$ are zero.
The relevant tight-binding parameters are presented in Table~\ref{table:multi_layer}.

\begin{table}[!h]
\begin{center}
\caption{Hopping parameters for 4-layer SRO$^{[1]}$}
\label{table:multi_layer}
\begin{tabular}{||c c c c c c c c c c||} 
 \hline
  &$t_{p,1}$(ev) & $t_{p,2}$(ev) & $t_{p,3}$(ev) & $t_{p,4}$(ev) & $U$(ev) & $u_{p,1}$(ev) & $u_{p,2}$(ev) & $f$(ev) & $\lambda$(ev) \\ [0.5ex] 
 \hline\hline
$p=1$ & 0.367 & -0.03 & 0.35 & 0.12 & 0.2 & 0.320 & 0.15 & -0.03 & 0.06 \\
 \hline
 $p=2$ & 0.348 & -0.03 & 0.35 & 0.12 & 0.2 & 0.253 & 0.15 & -0.03 & 0.06 \\
 \hline
 $p=3$ & 0.290 & -0.03 & 0.30 & 0.10 & 0.2 & 0.213 & 0.15 & -0.03 & 0.06 \\
 \hline
 $p=4$ & 0.250 & -0.03 & 0.25 & 0.01 &     &       &      & -0.03 & 0.06 \\ [1ex] 
 \hline
\end{tabular}
\end{center}
\end{table}

The band dispersion of the 4-layer SrRuO$_3$ film is plotted in Fig.~\ref{fig:Fig5}(b).
Multiple partial NFBs can be observed along the $\Gamma$X and XM directions. 
Since each NFB is doubly degenerate, there are eight partial NFBs along each direction. 
Although the spin-orbit coupling (SOC) slightly gaps out the NFBs at several points, the guidelines indicate that these bands were originally NFBs before the SOC was introduced. 
We note that the origin of the NFBs is from the appearance of EFBs in the model with $t_{p,1}=t_{p,2}=f=\lambda =0$, as plotted in Fig.~\ref{fig:Fig5}(c).
Let us denote this unperturbed EFB-model by $H^{(0)}_{\mathrm{4-layer}}(\mathbf{k})$.
The energies of the EFBs are given by 
\begin{align}
    E^{(0,n,\pm)}_{\mathrm{4-layer}}(k_x,0) &= \pm\sum_{p=1}^{3} 2(\tilde{f}_{p,n}^+\tilde{f}_{p+1,n}^+)(u_{p,1}+u_{p,2}),\\
    E^{(0,n,\pm)}_{\mathrm{4-layer}}(\pi,k_y) &= \pm \sum_{p=1}^{3} 2(\tilde{f}_{p,n}^-\tilde{f}_{p+1,n}^-)(u_{p,1}-u_{p,2}),
\end{align}
where $n$ is 0 or 1.
Here, $\tilde{f}_{p,n}^\pm = \tilde{f}_{p,n}^\pm/(\sum_{q=1}^4 |f_{p,n}^\pm |^2)^{1/2}$, where
\begin{align}
    f_{1,n}^{\pm} &= (\sqrt{b_n^{\pm}}(b_n^{\pm}-2(a_2^{\pm})^2-2(a_3^{\pm})^2))/(2\sqrt{2}a_1^{\pm}a_2^{\pm}a_3^{\pm}),\\
    f_{2,n}^{\pm} &= (b_n^{\pm}-2(a_3^{\pm})^2)/(2a_2^{\pm}a_3^{\pm}),\\
    f_{3,n}^{\pm} &= \sqrt{b_n^{\pm}}/(\sqrt{2}a_3^{\pm}),\\
    f_{4,n}^{\pm} &= 1,
\end{align}
with
\begin{align}
    b_n^{\pm} &= (a_1^{\pm})^2 + (a_2^{\pm})^2 + (a_3^{\pm})^2 +(-1)^n\sqrt{[(a_2^{\pm})^2 + (a_1^{\pm} + a_3^{\pm})^2][(a_2^{\pm})^2 + (a_1^{\pm} - a_3^{\pm})^2]},\\
    a_p^{\pm} &= u_{p,1} \pm u_{p,2}.
\end{align}
%
%
%
%
\( f_{a,n}^{\pm} \) represents the amplitude of the \( d_{yz} \) and \( d_{xz} \) orbitals in the \( a \)-th layer of the four partial EFBs eigenvectors labelled by $n$ and $\pm1$. 
The corresponding unnormalized eigenvectors corresponding to the partial EFBs along $\Gamma$X and XM are given by

\begin{align}
    |\mathbf{v}^{(0,n,\pm)}_{\mathrm{4-layer},\sigma}(k_x,0)\rangle &= \sum_{a=1}^{4}(\pm1)^af_{a,n}^+|d_{a,yz},(k_x,0),\sigma\rangle,\\
    |\mathbf{v}^{(0,n,\pm)}_{\mathrm{4-layer},\sigma}(\pi,k_y)\rangle &= \sum_{a=1}^{4}(\pm1)^af_{a,n}^-|d_{a,xz},(\pi,k_y),\sigma\rangle,
\end{align}
respectively.
%
%
In this EFBs model with \( t_{p,1} = t_{p,2} = f = \lambda = 0 \), the interlayer hopping processes exist only between \( d_{yz} \) and \( d_{xz} \) orbitals.  
One can construct the CLSs corresponding to the partial EFBs from $|\mathbf{v}^{(0,n,\pm)}_{\mathrm{4-layer},\sigma}(k_x,0)\rangle$ and $|\mathbf{v}^{(0,n,\pm)}_{\mathrm{4-layer},\sigma}(\pi,k_y)\rangle$.
The CLS obtained from $|\mathbf{v}^{(0,n,\pm)}_{\mathrm{4-layer},\sigma}(k_x,0)\rangle$ is extended along the $y$-axis while compactly localized along the $x$-axis.
On the other hand, the CLS extracted from $|\mathbf{v}^{(0,n,\pm)}_{\mathrm{4-layer},\sigma}(\pi,k_y)\rangle$ is elongated along the $x$-axis and compactly localized along the $y$-axis.

We apply the first-order perturbation theory to investigate how the partial EFBs deform by turning on \( t_{p,1} \) and \( t_{p,2} \), whose values are given in Table~\ref {table:multi_layer}.
The perturbative parts of the band dispersions are evaluated as
\begin{align}
    E^{(1)}_{\mathrm{4-layer}}(k_x,0) &= \sum_{p=1}^{4}(\tilde{f}_{p,n}^+)^2(t_{p,2}\cos{k_x}+t_{p,1}\cos{k_y})=\sum_{p=1}^{4}(\tilde{f}_{p,n}^+)^2(t_{p,1}+t_{p,2}\cos{k_x}),\label{eq:4_layer_cls_1}\\
    E^{(1)}_{\mathrm{4-layer}}(\pi,k_y)&=\sum_{p=1}^{4}(\tilde{f}_{p,n}^-)^2(t_{p,1}\cos{k_x}+t_{p,2}\cos{k_y})=\sum_{p=1}^{4}(\tilde{f}_{p,n}^-)^2(-t_{p,1}+t_{p,2}\cos{k_y}),\label{eq:4_layer_cls_2}
\end{align}
along $\Gamma$X and XM, respectively.
Interestingly, despite the relatively large value of \( t_{p,1} \), the bands remain nearly flat.
This is because the hopping corresponding to \( t_{p,1} \) does not break the stabilization of the CLSs obtained from $|\mathbf{v}^{(0,n,\pm)}_{\mathrm{4-layer},\sigma}(k_x,0)\rangle$ and $|\mathbf{v}^{(0,n,\pm)}_{\mathrm{4-layer},\sigma}(\pi,k_y)\rangle$.
This simply leads to the energy shifts of these CLSs as shown by the constant terms in (\ref{eq:4_layer_cls_1}) and (\ref{eq:4_layer_cls_2}).
While the dispersions are solely determined by the \( t_{p,2} \), this value is smaller than \( t_{p,1} \) by an order of magnitude.
This is why we have nearly flat bands, although \( t_{p,1} \) is not negligible.
These perturbative energy spectra agree well with the results shown in Fig.~\ref{fig:Fig5}(d), where only $t_{p,1}$ and $t_{p,2}$ are turned on, while $f$ and $\lambda$ remain zero.
Furthermore, even when $f$ and $\lambda$ are subsequently included, the perturbative spectra continue to closely match the exact band structure.

\section{Discussion}
We have analyzed the origin of the partial NFBs of the monolayer and 4-layer SrRuO$_3$ films from the perspective of the stabilization of the stripe-type CLSs via destructive interference.
In the monolayer case, we have demonstrated that one can engineer the CLS localized along all directions by arranging vacancies periodically at Ru sites.
This arrangement allows for the formation of a complete EFB, significantly enhancing the van Hove singularity at the energy corresponding to the complete flat band.
Although flat bands have not been widely observed in oxide materials, our results demonstrate that vacancy engineering can artificially induce flat bands in these materials.
Therefore, oxide materials may exhibit a broader range of emergent many-body phenomena driven by the enhanced many-body instability from the flat bands.
While Ru vacancies are common in SRO systems, achieving their periodic arrangement remains a challenge. However, it is known that scanning tunneling microscope (STM) tips can be employed to create vacancies in a controlled manner~\cite{jansen2024tip}. Moreover, in SRO films, various surface reconstructions can be induced through different growth processes~\cite{ohsawa2014visualizing,shin2010oxygen}, hinting that similar treatments might enable the reordering of vacancies.
%
%
Rather than creating a vacancy at a Ru site, a similar effect can be achieved by altering the conditions of the Ru site.
For example, in the biphenylene network, fluorinating a carbon atom was shown to have an effect nearly equivalent to removing the carbon atom~\cite{mo2024engineering}. 
The $p_z$ orbital of the fluorinated carbon atom has an on-site energy distinct from that of the unfluorinated carbon, ensuring it does not influence low-energy physics.
In the case of the SrRuO$_3$ film, a similar effect is expected to be achieved by removing oxygen atoms surrounding the Ru atom or by utilizing a sublattice that aligns with the vacancy configuration.


\section{Acknowledgments}
\begin{acknowledgments}
J.W.R. was supported by the National Research Foundation of Korea (NRF) Grant funded by the Korean government (MSIT) (Grant nos. 2021R1A2C1010572 and 2022M3H3A1063074). H.L. was supported by the National Research Foundation of Korea (NRF) grant funded by the Korea government (MSIT) (No. RS-2025-00556701). J.W.R. and H.L. were supported by Global - Learning \& Academic research institution for Master’s·PhD students, and Postdocs (G-LAMP) Program of the National Research Foundation of Korea (NRF) grant funded by the Ministry of Education (No. RS-2023-00285390). This work is also supported by the Ajou University research fund.
\end{acknowledgments}

\section{Competing Interests}
The authors declare no competing interests.

\end{document}